\documentclass[12pt,fleqn]{article}[1994/01/01]
\usepackage{graphics}
\usepackage{psfig}
\usepackage{epsfig}
\usepackage{espcrc1}

\def\half{\mbox{\small{$\frac{1}{2}$}}}

\def\beq{\begin{equation}}
\def\eeq{\end{equation}}
\def\bea{\begin{eqnarray}}
\def\eea{\end{eqnarray}}

\def\eqlab#1{\label{eq:#1}}

\def\eqref#1{eq.~(\ref{eq:#1})}
\def\eeqref#1{(\ref{eq:#1})}

\def\sla#1{#1 \hspace{-2.4mm} \slash}

\def\al{\alpha}

\def\ga{\gamma} 
 \def\De{{\it\Delta}}

 \def\La{{\it\Lambda}} 

\def\si{\sigma} \def\Si{{\it\Sigma}}
   
\def\w{\omega}

\def\dd{{\rm d}}

\def\eg{{\it e.g.\ }}

\def\CF#1#2#3#4{#1 {#2} (#3) #4}  

\def\np {Nucl.~Phys.}
\def\prc {Phys.~Rev.~C}
\def\prd {Phys.~Rev.~D}

\def\rk{{\rm k}}

\def\N{{\rm N}}

\newsavebox{\poten}
\sbox{\poten}{\parbox{18cm}{
\resizebox{18cm}{!}{\includegraphics*[1cm,21.5cm][28cm,24.5cm]{gph4.ps} }} }

\title{ \vspace{-1.5cm} A relativistic dynamical model of $\pi N$ 
scattering
\thanks{ Based on a talk given at
the 15th Int.\ Conf.\ on Fewbody Problems in Physics (Groningen,
22-26 July, 1997), to appear in Proceedings (Nuclear Physics A, special issue).}
 }

\author{{V. Pascalutsa and J. A. Tjon}\\[3mm]
Institute for Theoretical Physics, University of Utrecht, \\
P.O. Box 80006, 3508 TA Utrecht, The Netherlands }

\begin{document}
\thispagestyle{plain}

\maketitle
\vspace{-4cm}\hfill THU-97/23 \vspace{3.5cm}

\begin{abstract}
We present a unitary relativistic quasi-potential model for
describing the low-energy $\pi N$ interaction, based on the equal time
Bethe-Salpeter equation. It preserves the covariant structure 
of a relativistic spin 1/2 particle for the nucleon propagator, 
to be contrasted to other quasi-potential approximations.
\end{abstract}

\section{INTRODUCTION}

In the description of dynamical models of the $\pi N$ system, 
the corresponding amplitude is determined as a solution of a 
scattering equation of the Bethe-Salpeter (BS) or 
Lippman-Schwinger type. The BS equation
in relativistic studies is usually approximated by a covariant
3-dimensional quasi-potential (QP) reduction.
In this framework one in principle has a lot of freedom, 
since there is no unique scheme for 
the choice of the equation. Certainly, one would like to restrict 
this freedom not only by fitting to experimental data,
some restrictions can come from various symmetries and 
consistency requirements, such as the correct low-energy limit,
the correct one-body limit of the equation \cite{Grs82}, 
etc.

Another  consistency requirement is the symmetry of the 
renormalization of the positive and negative energy-states.
We find that most of the relativistic QP equations used in practice
do {\it not} satisfy this requirement, suggesting that
relativistic covariance is violated in these QP formulations.
One particular choice however, the {\it equal-time} (ET) 
(or instantaneous) {\it approximation}\cite{tlh}
of the BS equation, does not suffer from this pathology.  

In the next section we consider  the self-energy calculation
and demonstrate how the equivalence of the positive and negative 
energy-state renormalization can be destroyed.
In section 3 we briefly present our model for pion-nucleon
scattering based on the solution of the BS equation
in the equal-time approximation. 

\section{SELF-ENERGY AND RENORMALIZATION}

Consider the dressed nucleon propagator given by
\beq
S(\sla{P}) = \left[ \sla{P} - m - \Si (\sla{P}) +i
\epsilon \right]^{-1},
\eeq   
where $\Si (\sla{P})$ is the self-energy.
Relativistic covariance under general Lorentz transformations
requires that
\beq
\eqlab{form}
\Si (\sla{P})= A(P^2) \  \sla{P}  + B(P^2).
\eeq
In the  c.m.\ frame $P=(P_0,\,\vec{0})$ the Dirac structure of
the self-energy simplifies to
$
 \Si(P_0)=\Si_+(P_0)\ga_+ + \Si_-(P_0) \ga_-, 
$
where $ \gamma_\pm=\half (I  \pm \ga_0)$.
A similar decomposition holds for the propagator,
\beq\eqlab{propcms}
S(P_0) = S^{(+)}(P_0)\ga_+ + S^{(-)}(P_0) \ga_-,
\eeq
with 
$S^{(\pm)}(P_0) = \pm [ P_0 \pm (- m - \Si_\pm (P_0) 
+i \epsilon)]^{-1}$.
Obviously, $S^{(+)}$ corresponds to the positive and $S^{(-)}$ to the
negative energy-state propagations. 
Since  \eqref{propcms} describes a relativistic spin 1/2 
propagation, it has to have poles at $P_0=\pm m$ with the same
field renormalization constant $Z_2$. This implies that 
near these poles we have to  satisfy the condition
\beq
\eqlab{posneg}
\Si_{r} (P_0) = \Si_{-r} (-P_0),\,\,\,\, r=\pm 1.
\eeq
It is easy to see that \eqref{posneg} is valid as long as
the self-energy can be written in the covariant form \eeqref{form}.
If \eqref{posneg} does not hold, the positive and negative 
energy-states  
have different renormalized masses and the standard renormalization 
procedure\cite{WGT96} breaks down.

Let us now study as a specific example the lowest order $\pi N$ 
bubble self-energy graph. This occurs in the dynamical models
under consideration. We have after
the partial-wave decomposition:  
\beq
\eqlab{self1}
\Si_{r}(P_0) = \frac{i}{\pi}
\int\limits_{-\infty}^\infty\!\frac{\dd k_0}{2\pi}
\int\limits_0^\infty\! \frac{\dd\rk}{2\pi}\,\rk^2\,
\sum_\rho G^{(\rho)}(\rk,k_0;P_0)\,\Phi_r^{(\rho)}(\rk,k_0;P_0),
\eeq 
where $G^{(\rho)}$ is the pion-nucleon propagator 
and $\Phi_r^{(\rho)}$ represents the  
$\pi NN$ vertex contributions with $\rho$ characterizing the
$\rho$-spin\cite{tlh} of the intermediate state.
Furthermore, the integration variable $k_0$ is the relative-energy 
variable, defined as
$k_0 = ( p_{\N 0}\,\hat{\w} -  p_{\pi 0}\,\hat{E})/P_0$,
where $p_{\N 0}$ ($p_{\pi 0}$) is the 0-th component of the
nucleon (pion) 4-momentum, and $\hat{\w} = (P_0^2 - m_\N^2 + m_\pi^2)/2P_0
= P_0 - \hat{E}.$

The singularities of $\Phi_r^{(\rho)}$ in the $k_0$ plane are 
associated with those of the potential and the form factors.
In the QP approximation, the $k_0$ is usually taken to be
constrained to a certain value, depending in general on the 
3-momentum ${\bf k}$.
Considering the spectator QP approximation \cite{GrS93}
(where only the contribution from either the pion or the nucleon
pole is taken), we find the somewhat surprising result that condition 
\eeqref{posneg} is not satisfied. The breakdown essentially amounts
to that the self-energy doesnot have the form \eeqref{form}. In
particular, 'scalar invariants' like $A$ and $B$ are in this
approximation functions of not only $P^2$, but also of $P_0$.
This signals also a violation of the charge
conjugation symmetry.
The same problem arises in the choices given by
Pearce and Jennings\cite{PeJ91}.

Removing the $k_0$ singularities (by taking $k_0=0$) in the  
$\pi NN$ vertex functions, we may perform the $k_0$-integration
explicitly by contour integration, leading essentially to the 
ET-approximation. One can readily demonstrate that this choice of 
QP prescription does satisfy the above discussed symmetry property. 

\section{A MODEL FOR THE $\pi N$ SCATTERING}

We turn now to the discussion of our model for pion-nucleon scattering.
We compute the amplitude by solving the BS equation 
in the ET approximation.
The potential we use is given by the tree diagrams in Figure 1.
\begin{figure}[h]
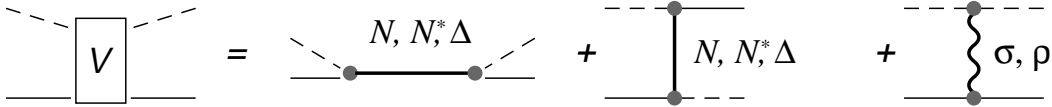

\vspace{-5mm}
\usebox{\poten}

\vspace{-7mm}
\caption{\sl
The model potential.}
\end{figure}
\vspace{-8mm}
 
We thus include
the $t$-channel $\si$ and $\rho$(770),
and the $s$- and $u$-channel $N$(938), $\De$(1232) and the Roper-resonance
exchanges. The $\De$ and the Roper are widthless when included
in the driving force, their one-pion-nucleon decay width is then
generated dynamically in the calculation. Note that, although the potential
is crossing symmetric, the kernel of the equation and thus the resulting
amplitude is not.

The bare vertices and propagators are
obtained from an effective Lagrangian of the meson and the isobar fields,
see \eg ref.\cite{PeJ91}. 
We allow the pion field to couple only through a derivative
coupling, which directly provides the correct low-energy limit, at least at
the tree level. The rescattering can in principle violate
the low-energy limit (in our model this may come due to 
the lack of the crossing symmetry). We have checked numerically that these
violations are small in our model.
\begin{table}[h]
\caption{ {\sl The model parameters which were ajusted to reproduce
the $\pi N$ phase-shifts.
Only the physical (renormalized) values of parameters are given.} }
\begin{tabular}{l|c}
\hline\hline
 {\rm coupling constants} & {\rm masses [GeV] }\\
\hline
$g_{\pi NN}^2/4\pi = 13.5$ & $\La_N = 1.5$, $\La_\pi=1.5$\\
$g_{\pi NN^*}^2/4\pi = 0.9$ & $\La_{N^*} = 1.9$, $m_{N^*}=1.54$ \\
$f_{\pi N\De}^2/4\pi = 0.33$, $z=-0.2$ & $\La_\De = 1.4$, $m_\De=1.24$\\
$g_{\si NN}g_{\si\pi\pi}/4\pi = 0.1$ & $\La_\si = 1.1$, $m_\si=0.55$\\
$g_{\rho NN}g_{\rho\pi\pi}/4\pi = 2.5, \kappa_\rho=3.7
$& $\La_\rho = 1.1$, $m_\rho=0.77$\\
\hline\hline
\end{tabular}
\end{table}
\vspace{-5mm}

For each particle we have used a form factor depending 
on the 4-momentum squared of the particle.
For a meson we take the one boson exchange form factor, and 
for a baryon we use the form factor of Pearce and Jennings\cite{PeJ91} (with 
$n_\al=2$).
The corresponding cutoff masses and other model parameters which
were fitted to the $\pi N$ phase-shifts are presented in Table 1.


For the propagator of the $\De$ we use the Rarita-Schwinger propagator,
and the $\pi N\De$ coupling is taken to be of the general structure
determined by a coupling  constant $f_{\pi N\De}$ and an off-shell parameter
$z$, cf.\ ref.\cite{NEK71}.
The $\rho$ exchange generates the conserved isovector-vector current.
The strength of its coupling is rather close to that determined
by the $\pi NN$ coupling
constant through the Kawarabayashi-Suzuki relation ( $g_{\pi NN}$ which
we use would imply $g_{\rho NN}g_{\rho\pi\pi}/4\pi \approx 2.8$). The $\rho$ 
thus mainly plays the role of the $\pi N$ contact term required
in the non-linear realizations of chiral symmetry. 
The $\si$ meson
is included so as to simulate the isoscalar-scalar contribution
of the correlated two-pion exchange. Therefore, for the $\si\pi\pi$
coupling constant we take the sign determined in ref.\cite{SDH94}, and we
find that this choice is preferable for the proper description of
the phase-shifts.  

Using this model we are able to get a reasonable fit to the elastic
$\pi N$ phase-shifts, see Figure 2, with the parameters given in
Table 1. To give a feeling about
the size of the rescattering contributions in our model, we also 
show (dashed lines) the results of the calculation where the 
principal part of the rescattering
integrals is neglected (the K-matrix approach).
\begin{figure}[h]
\vspace{-5mm}
\epsfysize=8.6cm
\epsfig{file=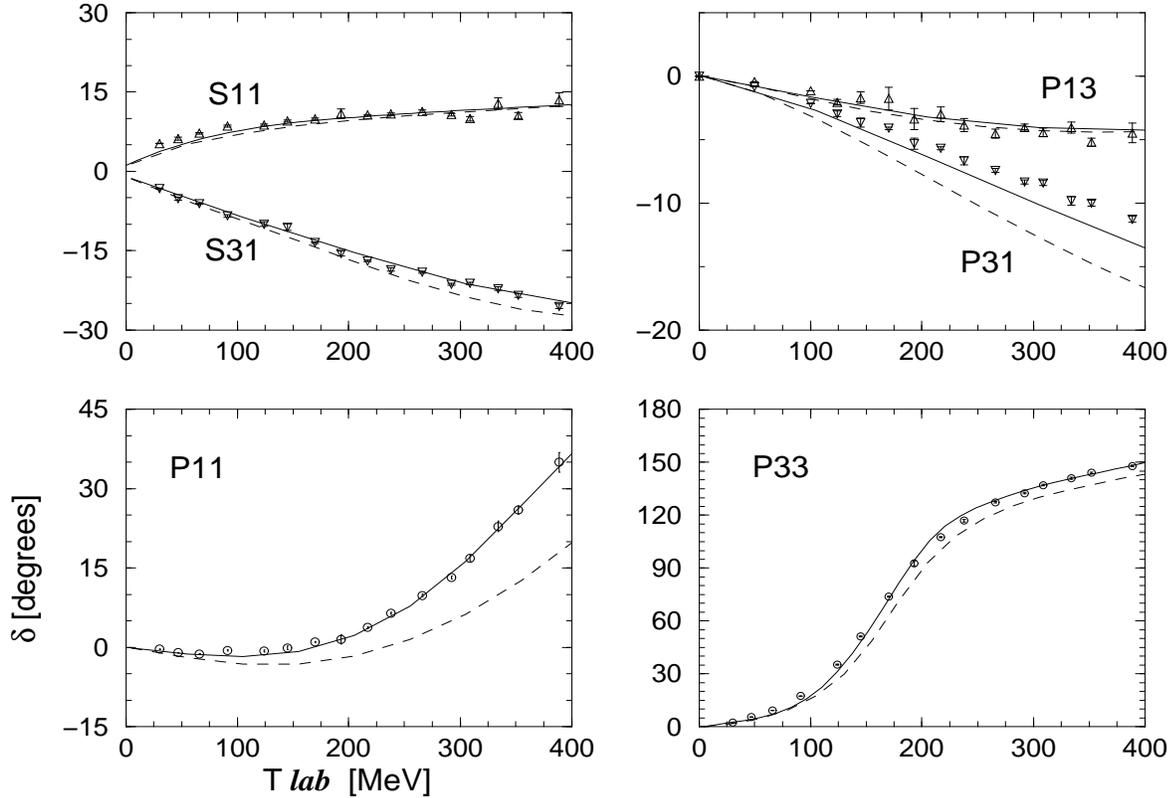}
\vskip7mm
\caption[F4]{ {\sl The S- and P-wave $\pi N$ phase-shifts. 
Solid lines 
represent the full calculation, dashed lines are the 
K-matrix predictions (with the same set of parameters).
Data points are from the SM95 partial-wave analysis\cite{ASW95}.} }
\end{figure}

\end{document}